 \documentclass[final,5p,times,twocolumn]{elsarticle}
 \biboptions{comma,sort&compress}
 \usepackage{epsfig}

\def\nin{\noindent}
\def\beq{\begin{equation}}
\def\eeq{\end{equation}}
\def\bea{\begin{eqnarray}}
\def\eea{\end{eqnarray}}

\def\oq{\char'134}
\def\as{\alpha_{\rm s}}
\def\amz{\as(M_{\rm Z^0})}

\journal{Nuc. Phys. (Proc. Suppl.)}

\begin{document}

\begin{frontmatter}



\title{Data Preservation in High Energy Physics -- 
  \\ why, how and when ?}

 \author[label1]{Siegfried Bethke \corref{cor1}
}
  \address[label1]{Max-Planck-Institut f\"ur Physik, 
\\
80805 Munich, Germany.}
\cortext[cor1]{Talk presented at QCD10, Montpellier, France, June 2010.}
\ead{bethke@mpp.mpg.de}



\begin{abstract}
\noindent
Long-term preservation of data and software of large experiments and 
detectors in high energy physics is of utmost importance to 
secure the heritage of (mostly unique) data and to allow advanced 
physics (re-)analyses at later times.
Summarising the work of an international study group,
motivation, use cases and technical details are given for an organised 
effort to secure and enable future use of past, present and future
experimental data. 
As a practical use case and motivation, the revival of JADE data 
and the corresponding latest results on measuring
$\as$ in NNLO QCD are reviewed.

\end{abstract}


\end{frontmatter}


\section{Introduction}
\nin
Analyses of data from large scale projects in experimental
particle physics are usually pursued for a typical time period of
5 years after close-down of the experiments.
After this time of post-mortem analyses, the number of active members of
large collaborations deteriorates to zero, as does the active
maintenance of data and software which is needed to efficiently
analyse these data.
While the data, as e.g. those obtained from 11 years of running of the 
electron-positron collider LEP, or from 16 years of running of the
lepton-hadron collider HERA, remain to be of unique importance and relevance
for the field of high energy particle physics, the long-term storage of
these data and - especially - the possibility to analyse these data
using the mandatory software packages and know-how of detector
particularities is, in almost all cases, not warranted after a relatively
short period of time.
In fact, at this time, the data of many past collider projects are
already lost, or are in an unusable state, and as this contribution is
being written up, data e.g. from LEP experiments continue to become lost
forever.

An international study group for data protection and future use
of high energy physics data, DPHEP, has formed and has presented 
\cite{dphep1} its first assessments of possible use cases and the technicalities of data and software preservation.
In the following, the physics case for data preservation and re-analysis
will be given, 
and will be demonstrated
by recent physics results obtained from  
data of the JADE experiment
which operated from 1979 to 1986 at the $e^+e^-$ collider PETRA
at DESY. 
Some details of preservation models will also be summarised.

\section{Physics case}
\nin
The most important scientific reasons for long-term 
data protection and future use
of data from past experiments are the following:

\begin{itemize}
\item
long term completion and extension of the scientific program of the project:
\\ The original program of a large scientific project is usually not completed at times of shutdown of the experiment(s), and is not completely finalised even after a period of a few additional years of data 
analyses, when availability and usability of data and software deteriorate 
due to the fast development of storage and computer hard- and software 
systems, and due to the fading availability of expert 
knowledge and personnel.
\item
cross collaboration analyses:
\\ maximum value and sensitivity of the data of large collider projects can
be achieved by combining the data statistics of several experiments which
operated on these facilities.
Such combinations, however, are often not completed, or not even started at the time when projects end.
\item
data re-use:
\\ due to the general development of scientific knowledge, new questions
may arise and/or new theoretical models and experimental methods may 
become available which make re-analyses of old data mandatory, if no such 
data are available from newer or active projects.
\item
training, education and outreach:
\\ there are many examples for successful use of data and analysis tools
from past experiments to train and educate students and even pupils on
modern scientific questions and methods; data, results and simulations of
past experiments are often used for outreach purposes since the 
\oq owner's rights" on public access to such data are usually less 
restrictive after close-down of the project.
\end{itemize}

\section{The JADE experiment at PETRA: past and presence}
\nin
One of the few practical examples of successful usage of data from a
large experiment, up to 30 years after the data have actually been taken, 
is the preservation and reanalysis of data from the JADE experiment
\cite{naroska} which operated at the PETRA $e¡+e¡-$ collider 
\cite{petra} at the DESY laboratory at Hamburg. 

JADE was one of the first symmetric and maximally hermetic multi-purpose, electronic particle detectors, comprising a high
resolution gas tracking (jet) chamber, placed in a hermetic solenoidal
magnetic field of 0.5 Tesla, surrounded by an electromagnetic calorimeter
and a hermetic muon filter and muon detector system.
PETRA delivered electron-positron collisions at centre of mass energies
from 12 to 46 GeV.
In total, about 200 $pb^{-1}$ of high quality collision data
was taken by JADE during its lifetime, corresponding to about 
45.000 well reconstructed multi-hadronic final state events ÷\cite{naroska}.

JADE, together with 4 other experiments at PETRA, took data from
1979 to 1986, when the PETRA collider was shut down and construction
work for the HERA collider began.
The data and software files continued to be
actively used for few more years, until 1990/1991, when the last
analysis results were published. 
After that time, the data, residing on archive tapes,
where physically removed
from the DESY computer centre and stored in aluminium boxes. 
Space limitations at DESY imposed physical destruction of these tapes 
by 1997.

The source code of the JADE software framework was collected and stored on
private computer accounts which were maintained until the IBM main frame
computer was phased out at DESY in 1997. 
The JADE collaboration had no
plan nor model for further data preservation and future use of their data.

The post-mortem project of JADE data and software revival is 
due to the interest and initiative of a few individual previous JADE
members, which started in 1995 to 1997, just in time to prevent inevitable
loss of data and software.

In 1997, about 1 TB of raw and calibrated data and MC production were moved
from a few thousand archived round tapes (160 MB per tape) to 600
IBM 3490 tapes (800 MB per tape).
A second copy of the data was made on 200 Exabyte cartridges 
(2.5 GB each) \cite{olsson}. 
No MC generated data files were preserved at that stage.
In 2005, the exabyte cartridges travelled to Munich, were 
transferred to disk, and are now a (very small) part of the ATLAS data 
storage at the LHC Tier-2 centre of the Max Planck Society computing 
centre at Garching.

The reactivation of the software 
was successfully completed in 1999 \cite{fernandez}. 
It required adapting the JADE software code, originally consisting 
of FORTRAN-IV, but also partly of SHELTRAN, MORTRAN and Assembler code,
to UNIX platforms and modern FORTRAN compilers.

Today, the generation of model collision events,
using modern physics Monte Carlo Generators with full detector
simulation, is possible again, and simulated as well as real data
events can be examined using a revived version of the original JADE event
display with enhanced options like colour (which was not available
during JADE running time).
The revived JADE software runs on IBM AIX machines, relying on
the fact that these systems utilise the same byte order as the IBM 370 did.
The revitalisation, details of emulation routines and the usage of the
software packets and data files is documented in a respective JADE 
computing note
\cite{fernandez-note}.

The complexity of the software code and the data structures, of the detector
hardware and its simulation, however, still requires the knowledge of 
experts for analysing these data. 
This knowledge is currently maintained at the Max-Planck-Institute for
Physics at Munich and at DESY.
The data and usage thereof is still \oq owned" by the original JADE
collaboration, such that no general \oq open access" to these data is granted.

\section{Physics benefits: new results from old data}
\nin
Improvements motivating reanalyses of the JADE data, due to
advanced theoretical knowledge and analysis methods compared to
those being available at PETRA times, are summarised, with special
attention to the study of hadronic final states, in Table~1.
Enhanced and more profound theoretical knowledge, more sophisticated 
Monte Carlo (MC) and hadronisation models, improved and optimised
experimental observables and methods, and a much deeper understanding and
precise knowledge of the Standard model of electroweak and strong
interactions make it mandatory and beneficial to reanalyse
old data and to significantly improve their scientific impact.

{\scriptsize
\begin{table*}[hbt]
\setlength{\tabcolsep}{1.4pc}
 \caption{\scriptsize    Possible improvements motivating reanalyses of
old data, with past and presently available knowledge and methods.}
    {\small
\begin{tabular}{lll}
&\\
\hline
improvement &then (at/after PETRA) & now (after LEP)    \\
\hline
new and improved theoretical calculations & QCD in (N)LO & QCD in 
resummed NNLO \\
new and improved MC models & fixed order (N)LO & NLLA \& NLO shower \\
new and optimised observables & event shapes: T, S, O,... & $B_w$, $B_t$,
$D_3$, Durham jets, ... \\
more complete knowledge of Standard Model & --- & top quark, W, Z, ... \\
\hline
\end{tabular}
}
\label{tab:benefits}
\end{table*}
}

In general, these benefits can be used to
\begin{itemize}
\item
re-do previous measurements, with increased precision and reduced 
systematic uncertainties;
\item
perform new measurements, at Energies and processes 
where no other data are available today;
\item
if new phenomena are found today: go back and check at lower energies.
\end{itemize}

\subsection{JADE data and LEP parametrisation:
universality of hadronisation}
\begin{figure}[hbt] 
\centerline{\includegraphics[width=5.8cm]{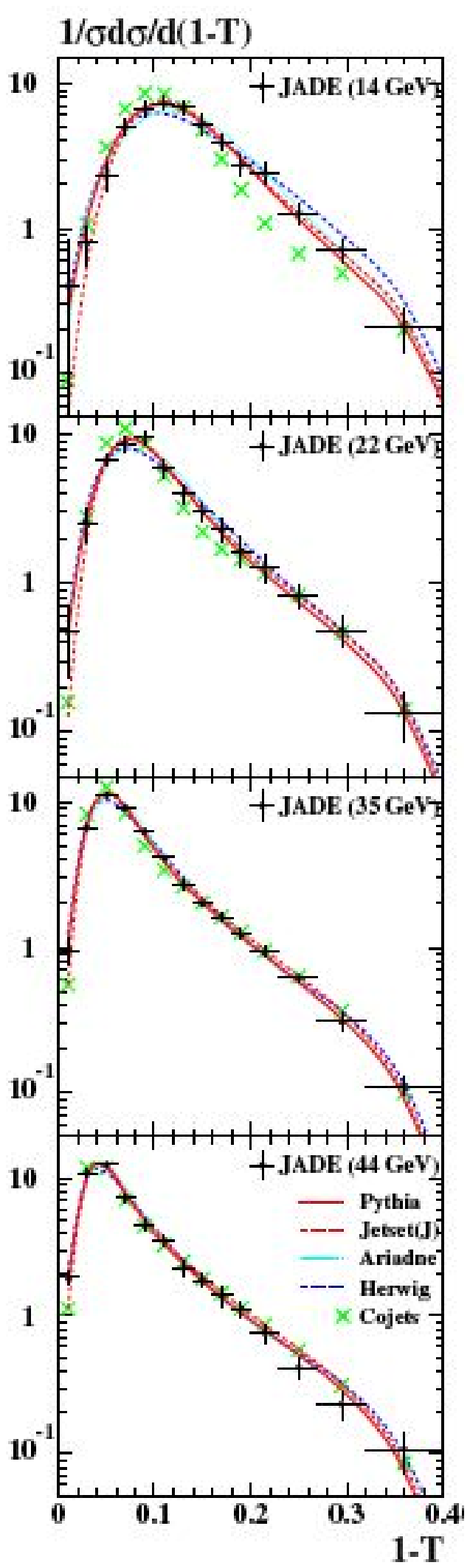}}
\caption{\scriptsize Distributions of 1-T data 
at c.m. energies from 14 to 44 GeV.
The data are compared to model predictions with
hadronisation parameters optimised at LEP.}
\label{1-t} 
\end{figure}

\nin
One of the first surprises when starting to reanalyse JADE data was
to realise that newly generated Monte Carlo model events, based
on modern QCD shower models like JETSET, HERWIG and PYTHIA, using
parameters as optimised much later by the experiments at the LEP
collider, described the JADE data, at much smaller c.m. energies than at
LEP, to a degree never obtained at PETRA times \cite{fernandez,kluth1}.
In detail, hadronic event shape distributions are correctly described
at all energies, down to 14 GeV, by the models without the need to
re-adjust model parameters at each c.m. energy 
(see Fig.~\ref{1-t}) - a fact never achieved
at PETRA times, where models required significant 
retuning of parameters at each major energy.

There is an important physics result behind this observation: 
the process of hadronisation as implemented in these models does not
depend on the c.m. energy, such that studies of physical parameters, like
the size and the energy dependence of the strong coupling constant, $\as$,
can be pursued with a minimum of systematic uncertainties.
Moreover, using the JADE data sample, such measurements can be done
in the entire PETRA energy range, where the energy dependence of $\as$
is expected to be much larger than at the higher energies of LEP.
Note that at PETRA times, the insufficient 
quality of modelling the
lowest energy data, around 14 and 22 GeV, prevented 
significant studies of those data.
%
\nin

\subsection{The running coupling $\as$ in NNLO QCD}
\nin
The latest study and re-use of JADE data \cite{jade-nnlo}
demonstrates the physical value
of old data at times far after the active time of the experiment:
measurements of the coupling parameter of the strong interaction, 
$\as$, can now be pursued with much higher precision and considerably
smaller systematic uncertainties than at the times of PETRA.
All of the facts listed in Table~1 apply and improve the significance
of such measurements today, using the data of the past.

The status of $\as$ measurements at the time of PETRA was reviewed
e.g. in \cite{altarelli,budapest}.
It can be summarised by quoting 
$\as(35~GeV) = 0.14 \pm 0.02$, 
where the error was largely dominated by hadronisation and QCD uncertainties.

\begin{figure}[hbt] 
\centerline{\includegraphics[width=7.cm]{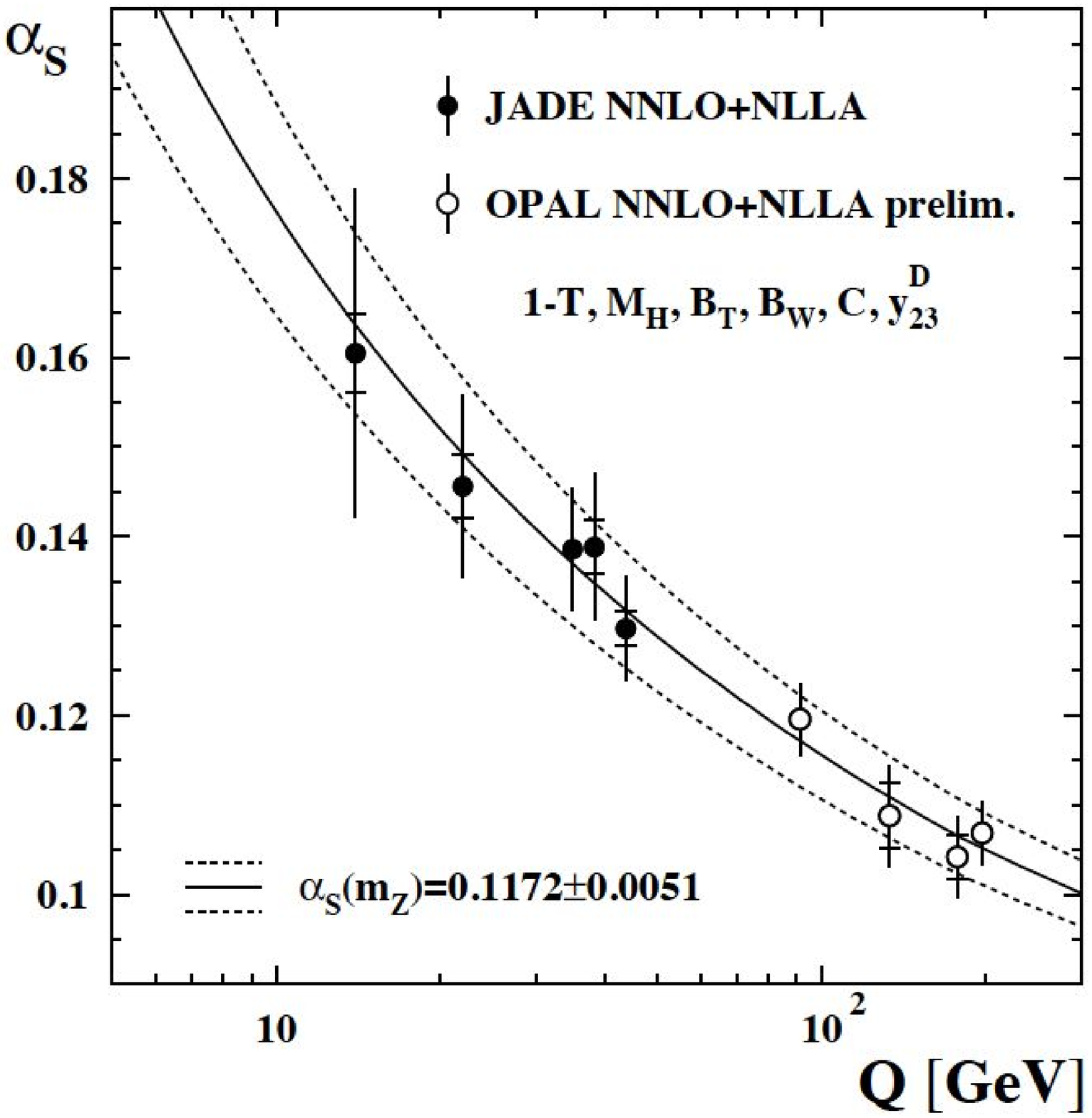}}
\caption{\scriptsize Measurements of $\as (Q)$ from JADE data,
in the energy range from $Q\ =\ $14 to 44 GeV, using event shapes and QCD predictions in resummed NNLO\&NNLA perurbation theory.
The results from a similar analysis of OPAL data (preliminary)
are also included.}
\label{as} 
\end{figure}

The results of reanalysing the JADE data \cite{jade-nnlo}, 
using modern event shape and jet distributions and the most recent and
advanced predictions in resummed next-next-to-leading order (NNLO)
of QCD perturbation theory \cite{nnlo+nlla}, are shown in
Figure~\ref{as}, as a function of the c.m. energy.
Also shown is the prediction of the running $\as$ in 3-loop QCD
perturbation theory, for the central fit value of $\amz$ to all JADE data,

$$\amz = 0.1172 \pm 0.0020 {\rm (exp.)} \pm 0.0046 {\rm (th.)}\ .$$

\nin
The results are also compared to a similar analysis using LEP data 
\cite{lep-nnlo-nlla}.

The value of reusing JADE data is obvious: $\as$ runs
with energy as predicted by QCD, which is significantly proven by the JADE
data alone, manifesting the concept of Asymptotic Freedom \cite{af}
(see \cite{as09} for a recent review of measurements of $\as$).
No other such results are currently available in this energy regime.
They are, due to many improvements in the field during the past 20 years,
significantly more precise than what has been achieved during and
shortly after the actual running of PETRA.
%

{\scriptsize
\begin{table*}[hbt]
\setlength{\tabcolsep}{1.4pc}
 \caption{\scriptsize    Data preservaion models and different use cases as
worked out by the DPHEP study group.}
    {\small
\begin{tabular}{lll}
&\\
\hline
{\bf Level} \ & {\bf Preservation model} & {\bf use case}    \\
\hline
1 & provide additional documentation & publication-related information search \\
2 & preserve the data in simplified format & outreach, simple training analyses \\
3 & preserve the analysis level software and data format & full scientific use based on existing reconstruction \\
4 &  preserve reconstruction and simulation software  & full potential of experimental data \\
\ & and basic level data & \ \\
\hline
\end{tabular}
}
\label{tab:models}
\end{table*}
}

\section{International effort of data preservation: DPHEP}
\nin
While the JADE example is one of the only existing examples of
preserving and reusing 
data and software of a complex experiment in high energy
physics, it is known that the data of many other experiments are
already lost inevitably, and/or cannot be used any more due to the
lack of functional software and analysis environment.
The LEP experiments, more than 10 years after active data taking, report
occasional analyses until today, however it is known that the data,
as well as the corresponding software environments, are beginning to
fade away, and some losses of data (archive tapes) 
were already communicated.

\begin{figure}[hbt] 
\centerline{\includegraphics[width=9.cm]{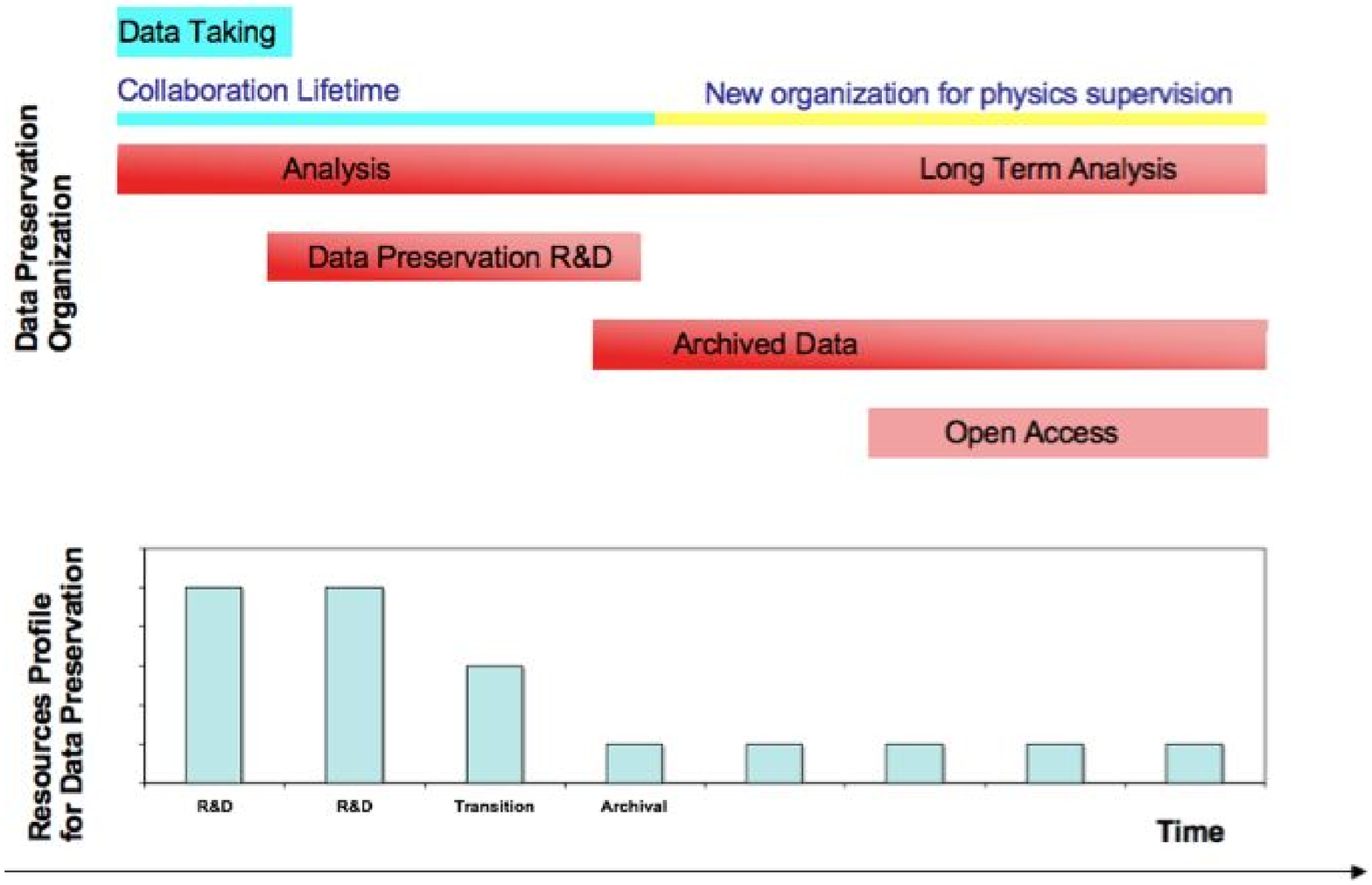}}
\caption{\scriptsize Timeline and need of resources
for a data preservation model 
at level 4, i.e. maintaining full flexibility for future 
and long-term use
of the data of a HEP experiment after its termination.}
\label{timeline} 
\end{figure} 

In 2009, an international initiative to preserve data in high energy
physics (DPHEP) has formed and worked out arguments, technical details and
governance policy for a concerted effort to preserve 
and re-use data of recent and current large-scale high energy physics
experiments.
In particular, 4 different levels of data preservation models have been
defined, as summarised in Table~\ref{timeline}.
These levels are inclusive, i.e. higher levels include the details
of those before.
In general, they differ in their overall purpose, in their degree of 
flexibility and in the amounts of efforts necessary to maintain these 
levels.

While levels 1 and 2 are realised in a number of projects and cases, they
do not allow to perform new or improved analyses (compared to what was 
published in the past).
Level 3 provides some limited means for new analyses.
Only level 4, however, gives the flexibility which 
provides full future potential of data analysis.
It is also the most intricate model, as it requires significant and
sustained efforts of preparation, maintenance and validation.

A typical time-line of a level-4 data preservation model is given
in Figure~\ref{timeline}.
It relates the times of data taking, of collaboration life time, 
of data preservation R\& D, of long term analysis and a possible 
and final \oq open access" period with the new organisation
of physics supervision and resources needed to pursue
such a project (in units of FTE's, as a function of the number of years).
Further details on the questions of technologies, facilities, funding, 
governance, supervision and authorship rights are elaborated and given
in \cite{dphep1}. 

Future usability and preservation of data of large HEP experiments is mandatory,
both on grounds of scientific importance and of sustainment of publicly 
funded heritage.  
While extra resources must be identified to pursue active data
preservation, the necessary amount of such resources is only at the level of very few percent of the original investments.
Failing to do so, i.e. accepting the loss of data and their scientific 
usability, may be regarded to be a crime, given the large 
amounts of expenses and
manpower which were invested in the original experimental programs.

\section*{Acknowledgements}
\nin
I would like to thank the LPTA-Montpellier and Stephan Narison for 
hospitality and for organising this pleasant conference.
I acknowledge the fruitful work of the members of the
DPHEP initiative.
Many thanks to P. Movilla Fernandez, J. Olsson, S. Kluth, J. Schieck, C. Pahl for their invaluable contributions to the JADE revival.


\end{document}